\documentclass[twocolumn,trackchanges]{aastex62} 

\usepackage{amsmath}
\graphicspath{{./}{figures/}}

\shorttitle{Warp evolution between 8$-$14 \,kpc}
\shortauthors{Wang et al.}

\begin{document}

\title{ Mapping the Galactic disk with the LAMOST and Gaia Red clump sample. VI:  An evidence for the long lived non-steady warp of non-gravitational scenarios} 
\author[0000-0001-8459-1036]{H.-F. Wang}
\affil{South$-$Western Institute for Astronomy  Research, Yunnan University, Kunming, 650500, P.\,R.\,China}
\affil{LAMOST Fellow }
\author{M. L\'opez-Corredoira}
\affil{Instituto de Astrof\'\i sica de Canarias, E-38205 La Laguna, Tenerife, Spain}
\affil{Departamento de Astrof\'\i sica, Universidad de La Laguna, E-38206 La Laguna, Tenerife, Spain}
\author{ Y. Huang}
\affil{South$-$Western Institute for Astronomy  Research, Yunnan University, Kunming, 650500, P.\,R.\,China}
\author{ J. Chang}
\affil{Purple Mountain Observatory, the Partner Group of MPI f$\ddot{u}$r Astronomie, 2 West Beijing Road, Nanjing 210008, China}
\author{ H.-W. Zhang}
\affil{Department of Astronomy, Peking University, Beijing 100871, People's Republic of China.}
\affil{Kavli Institute for Astronomy and Astrophysics, Peking University, Beijing 100871, People's Republic of China}
\author{J. L. Carlin}
\affil{LSST, 950 North Cherry Avenue, Tucson, AZ 85719, USA}
\author{ X.-D. Chen}
\affil{Key Laboratory of Optical Astronomy, National Astronomical Observatories, Chinese Academy of Sciences, Beijing 100012, People's Republic of China}
\affil{Department of Astronomy, China West Normal University, Nanchong 637009, China}
\author{ \v{Z}. Chrob\'{a}kov\'{a}}
\affil{Instituto de Astrof\'\i sica de Canarias, E-38205 La Laguna, Tenerife, Spain}
\affil{Departamento de Astrof\'\i sica, Universidad de La Laguna, E-38206 La Laguna, Tenerife, Spain}
\author{ B.-Q. Chen}
\affil{South$-$Western Institute for Astronomy  Research, Yunnan University, Kunming, 650500, P.\,R.\,China}
\correspondingauthor{HFW}
\email{ hfwang@bao.ac.cn{(\rm HFW)}};\\

\begin{abstract}

By combining LAMOST DR4 and Gaia DR2 common red clump stars with age and proper motion, we analyze the amplitude evolution of the stellar warp independently of any assumption with a simple model. The greatest height of the warp disk increases with Galactocentric distance in different populations and it is dependent on the age: the younger stellar populations exhibit stronger warp features than the old ones, accompanied with the warp amplitude $\gamma ({\rm age})$ decreasing with age and its first derivative $\dot{\gamma} ({\rm age})$ is different from zero. The azimuth of line of nodes  $\phi _w$ is stable at $-$5 degree without clear time evolution, which perfectly confirms some of previous works. All these self-consistent evidences support that our Galactic warp should most likely be a long lived, but non-steady structure and not a transient one, which is supporting the warp is originated from gas infall onto the disk or other hypotheses that suppose that the warp mainly affects to the gas, and consequently younger populations tracing the gas are stronger than the older ones. In other words, the Galactic warp is induced by the non-gravitational interaction over the disk models.

\end{abstract}

\keywords{Galaxy: kinematics and dynamics - Galaxy: disk - Galaxy: structure}

\section{Introduction} 

Most spiral galaxies have a warped disk \citep{Reshetnikov1998, Sanchez-Saavedra1990,San03}, although in some spirals warps cannot be observed due to their low inclination. Like many spiral galaxies in the universe, Milky Way's (MW) warp was detected by neutral hydrogen (HI) gas many years ago \citep{Kerr1957, Bosma1981, Briggs1990, Lev06b, Nakanishi2003}. Dust has also been observed in \citet{Marshall2006}. Furthermore, there are several works that measure parameters of the stellar Galactic warp from density maps/star counts \citep{Lop02b, Momany2006, Reyle2009, Amo17}.
There is also an intuitive 3 dimensional (3D) map of the Galactic warp traced by classical Cepheids \citep{Chen2019, Skowron2019}, 
including precession measurements consistent with the Brigg's rule \citep{Briggs1990} and showing us that its  amplitude of northern part is very prominent and stronger than that of the southern part \citep{Skowron2019b}. Kinematic signatures of the Galactic warp were studied by \citet{Huangyang2018, Schonrich2017} in the vertical velocity, angular momentum, azimuthal velocity or guiding center radius parameters space, showing that the Galactic warp is not a transient structure, which is consistent with the simple calculation of \citet{wang2020a}. More kinematical signals could also be found in \citet{Smart1998, Poggio2018, Gomez2019}, etc. The clear differences between thin disk and thick disk warp classified by metallicity and abundance is shown in one of our series of works \citep{Li2020}.

The mechanisms of formation of the gas and stellar warp were proposed in many works. \citet{Debattista1999, Shen2006} showed that warps were produced by the dynamical friction between a misaligned rotating halo and disk. In some cases, this misalignment might be related to misaligned gas infall \citep{Ostriker1989, Quinn1992, Bailin2003b}. \citet{Burke1957, Weinberg2006} suggested it was caused by the interaction with the Magellanic Clouds and \citet{Bailin2003a} proposed the cause of interaction with Sagittarius \citep{Bailin2003a}; \citet{Hunter1969} claimed that Magellanic Clouds mass is not enough to explain the observed amplitude of the warp, however, new elements of amplification have been introduced by \citet{Weinberg2006}. Other scenarios, that is to say, perturbations by dwarf satellites \citep{Weinberg1995, Shen2006}, intergalactic magnetic fields influence \citep{Bat98} or accretion/infall of the intergalactic medium flows directly onto the disk \citep{Lop02a}, disk bending instabilities by \citet{Revaz2004}, etc., also appeared. These and other mechanisms were proposed but some observational evidences may favour some of them, the different interpretations on the formation of the warp are still being hotly debated. In any case, we know the kinematical distributions of vertical bulk motions will be contributed by warp asymmetrical structure \citep{wang2018a, wang2020a}, in return, vertical motions can be used to constrain the warp's properties.

Warp in the Milky Way bends up upwards and downwards in the north and south hemisphere separately with different amplitude at least in the gas \citep{Lev06}. The amplitude of the warp clearly increases strongly with radius and varies with the azimuth angle \citep{Li2020, Liu173, Lop141}. A linear simple relation between the amplitude of the warp and the Galactocentric distance was used to explain the increase trend of vertical velocity with vertical angular momentum \citep{Schonrich2017}. In \citet{Lop141}, the vertical bulk motions are contributed by a warp with modeling as a set of circular rings  that are rotated and whose orbit is in a plane with angle with respect to the Galactic plane, with vertical amplitude $z_w(R,\phi) = \gamma (R $-$ R_{\odot})^\alpha \sin(\phi $-$ \phi _w)$, where $R$ is the Galactocentric distance and $\phi $ is the Galactocentric azimuth. With the calculation of $\gamma$, $\dot{\gamma}$ by assuming $\phi_w$ and $\alpha $ are constant, their work indicated that most likely the main S-shaped structure of the warp is a long-lived feature.

As mentioned in \citet{Lop141}, the precision of vertical velocity can be increased by at least an order of magnitude with the help of the Gaia proper motion \citep{Gaia2018, Katz2018} together with spectroscopically classified red clump stars, e.g. LAMOST \citep{cui2012, Deng2012, liu2014, zhao2012}. In addition to the unprecedented proper motion, we have stars age nowadays, so we have this first chance to research the evolution of the warp structure properties. For this work, we are motivated to use \citet{Lop141} simple model and LAMOST (Large Sky Area Multi-Object Fibre Spectroscopic Telescope) DR4, and Gaia DR2 common red clump stars to investigate the warp amplitude, its first derivative and  greatest height variation with different age populations so that we could get better constraint for the $\gamma$ , $\dot{\gamma}$, and $\phi_{w}$. Hence, we could offer our interpretation of the formation and evolution history of Milky Way warp.

This paper is structured as follows: The section 2  is about how we select our red clump stars, velocity derivation, and the vertical velocity distribution in different age populations. Model and method are introduced in the section 3 . Our results will be shown in section 4 and discussions are displayed in the section 5. Finally we will give our conclusions of this work.

\section{The Sample Selection}  

During this work, we use the red clump giants  selected  from the LAMOST Galactic spectroscopic surveys and Gaia astrometric survey. The scientific motivations and target selections of LAMOST phase I can be found in \citet{cui2012, Deng2012, liu2014, zhao2012}. Now, we are entering into the Phase-II. Its fiber is 3.3 arcsec and the mean seeing during LAMOST observations is around 3 arcsec  and the spatial resolution of LAMOST should be around 5 arcsec. Selection functions of LAMOST is almost a flat along apparent magnitude, more details can also be found in \citet{Carlin2012, Yuan2015, Liu171}. We select stars in LAMOST DR4, it has 3,461 observed plates and stellar parameters for 6,597,527 spectra are derived in \citet{Xiang2017c}. Total sample size is 7,620,612 including stars and galaxies.

The red clump stars selection details can be found in \citet{Huang2020, Huangyang2015}, the distance and age is determined by the Kernel Principal Component Analysis (KPCA) method, which could be found more details in \citet{Xiang2017b, Xiang2017c}. According to \citet{Huang2020}, the distance uncertainties are around 5-10\% and age uncertainties are around 30\%, which have been used quite well in \citet{wang2019, wang2020a, wang2020b}, and red clump stars are well known horizontal stars and standard candles so that it is not strange that the distance error of our sample is small. As described in \citet{Huang2020}, with the help of positions in the metallicity$-$ dependent effective temperature$-$surface gravity and color$-$metallicity diagrams, red clump stars could be selected. Using supervised by the high-quality asteroseismology data and Kernel Principal Component Analysis (KPCA) method, ages are determined. Using the properties of intrinsic absolution magnitude, extinction with star pairs method \citep{Yuan2013}, re-calibration of the Ks absolute magnitudes considering both the metallicity and age dependences, we could acquire the distance with uncertainties of 5$-$10 percent. The Gaia DR2 catalogue contains high-precision positions, parallaxes, and proper motions for 1.3 billion sources as well as line-of-sight velocities for 7.2 million stars. For stars of G $ < $ 14 \,mag, the median uncertainty  is 0.03 \,mas for the parallax and 0.07~mas~yr$^{-1}$ for the proper motions. In order to get reliable stellar parameters and try to reduce the halo contamination from the 0.14 million sample mainly consisted of primary red clumps (RCG) with few contaminations of secondary red clumps and Red Giant Branch(RGB) stars, The typical purity and completeness of our primary RC sample are all greater than 80 per cent and we focus on the kinematics but not the star counts so the completeness may have very minor influence on the study. We use the latest sample of this catalog to investigate our scientific target carefully according to the following criterions:

\begin{enumerate}
\item Sample without parameters such as distance, radial velocity, temperature and surface gravity are removed.  

\item Stars located inside $|Z|$ $<$ 1 kpc and 8 $<$ R $<$ 14 \,kpc are chosen. 

\item Stars with LAMOST spectroscopic SNR $>$ 20 and age less than 15 \,Gyr are included. 

\item $[Fe/H]$  $\textgreater$ $-$1.3 \,dex.

\item $V_R$=[-150, 150] km s$^{-1}$, $V_{\theta}$=[-50, 350] km s$^{-1}$, and $V_Z$=[-150, 150] km s$^{-1}$.
\end{enumerate}

\begin{figure*}
  \centering
  \includegraphics[width=0.98\textwidth]{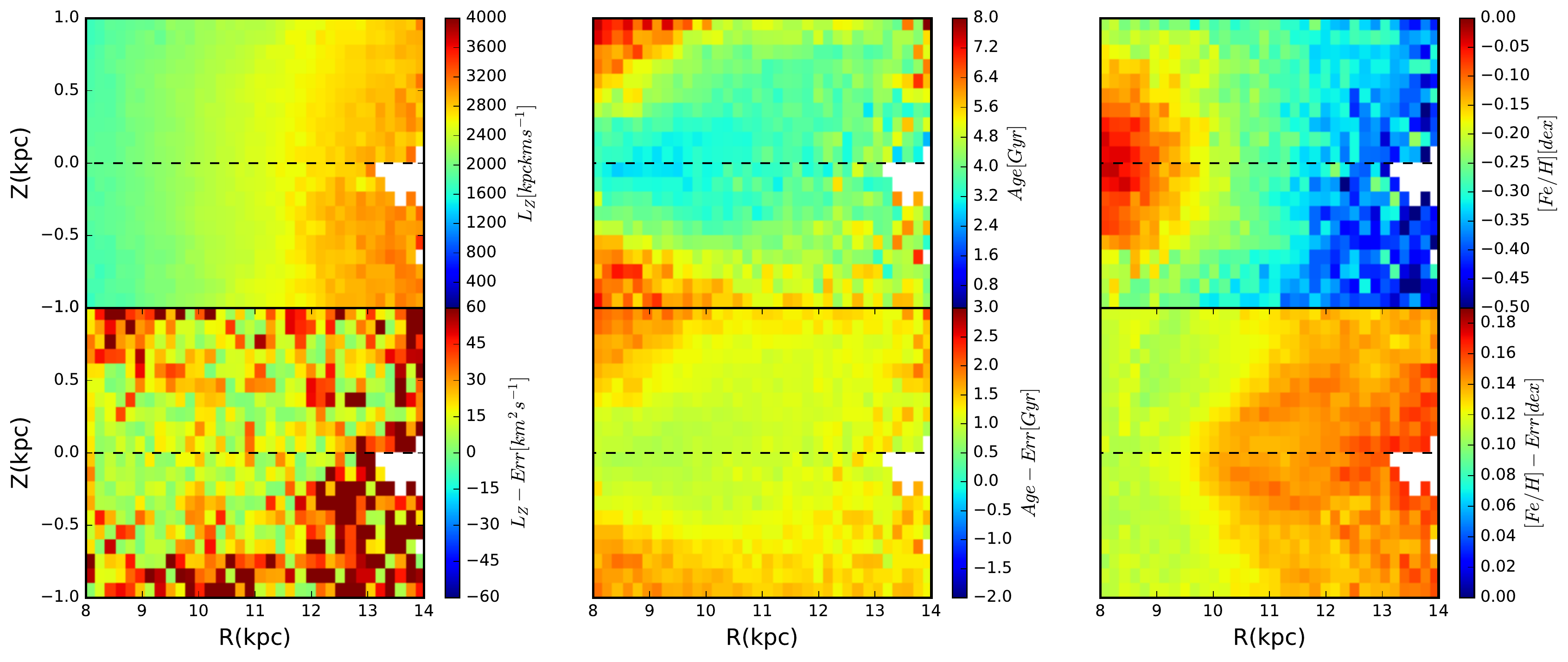}
  \caption{The top left figure shows the angular momentum of the disk distribution in the $R, Z$ plane, the errors of these stars are bootstrap error. The derived age and corresponding measured age error of the method are shown in the middle panel in the $R, Z$ plane. The last one is the metallicity and its measured error of the method distribution in the $R, Z$ plane. All these stars are limited in the range of $R$=[8 14] \,kpc, $Z$=[-1 1] \,kpc and the minimum number of stars in every pixel is five that is enough for us  to see the general pattern of data.} 
  \label{warp_RZXYagedisLz}
\end{figure*}

We derive the 3D velocities assuming the location of Sun is $R_{\odot}$  = 8.34 \,kpc \citep{Reid14} and $Z_{\odot}$ = 27 \,pc \citep{Chen01}, \citet{Tian15} solar motion values: [$U_{\odot}$, $V_{\odot}$, $W_{\odot}$] = [9.58, 10.52, 7.01] km s$^{-1} $, other solar motions \citep[e.g., ][]{Huang2015} won't change our conclusion at all. The circular speed of the LSR is adopted as 238 km s$^{-1}$ \citep{Schonrich12}. and Cartesian coordinates on the basis of coordinate transformation described in $Galpy$ \citep{Bovy2015} with LAMOST radial velocity with the precision better than 5 km s$^{-1}$, which is  more convenient and direct than those described on by one in \citet{wang2018a, wang2019, wang2020a, wang2020b}. The vertical angular momentum distributions in per unit mass of the sample associated with the error analysis in  the $R, Z$ plane of the Cartesian coordinate system are shown in left panel of  Fig. \ref{warp_RZXYagedisLz}. Meanwhile, the projected age, $[Fe/H]$ and its measured error distributions are displayed in the middle and right panel respectively. It reasonably implies that the angular momentum of stars increases with radial distance in the disk, including the corresponding errors, due to that disk total momentum is approximate to be vertical momentum. The age distribution for which the stars inside are relatively older and those outside are generally younger support the  inside$-$out formation of the Milky Way disk, age error is smaller comparing with the age value, albeit probably some possible systematic errors that remain to be ignored, it can still give us the good chance to map the dynamical structures in different age populations. 

As shown in the bottom right one of Fig \ref{warp_RZXYagedisLz}, with Galactic radial distance increasing, the metallicity has a negative trend and the error is also becoming larger with value of around 0.1$-$0.15 \,dex, which is reasonable. Here we want to emphasize that, by using this recent updated sample, some asymmetrical structures such as radial or bulk motions reconstructed here are very similar to our previous series of works about the Galactoseismology \citep{wang2020a} and recent Gaia 3-dimensional kinematics works \citep{Katz2018,Lop19}. During current work, we focus on the range of $R$=[8 14] \,kpc, $Z$=[-1 1] \,kpc and the bins with minimum number of every pixel containing five stars are shown in Fig \ref{warp_RZXYagedisLz}.

\begin{figure*}
  \centering
  \includegraphics[width=0.78\textwidth]{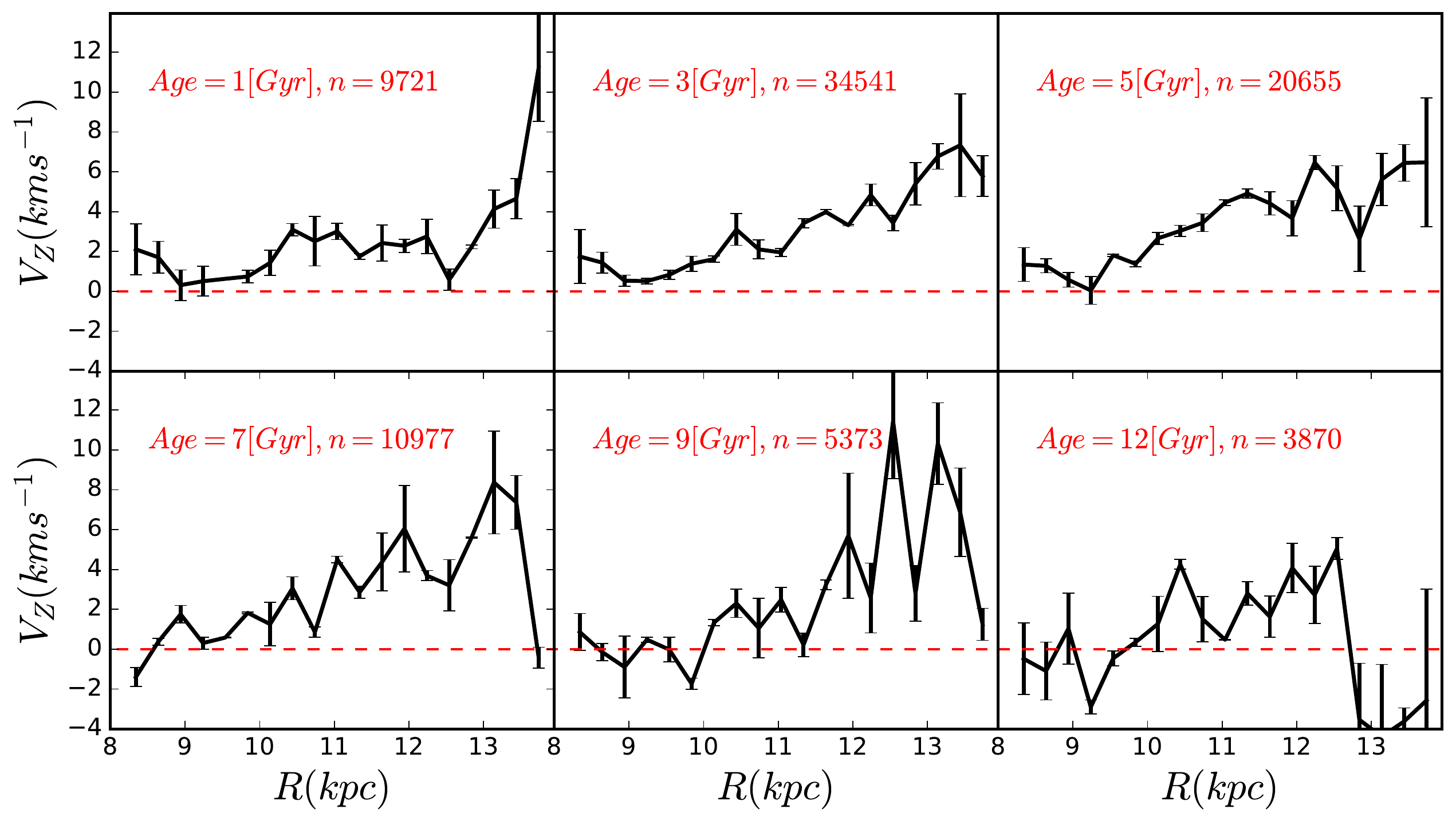}
  \caption{Vertical velocity distribution with radial distance in different age populations. From the top left to bottom right, corresponding to [0 2], [2 4], [4 6], [6 8], [8-10], [10 14] \,Gyr. Note that each R bins for every age population contains at least 20 stars and the star number and median value of each population are labeled in the top left of each panel by red color. Almost all populations increase with distance for the overall trend, although there are  some oscillations. The horizontal red dashed line is the zero of velocity value, used to guide our eyes.}
  \label{warp_Rvz}
\end{figure*}

We could see the vertical velocity ($V_{Z}$) distribution of our sample with radial distance ($R$) in different age populations with different stellar ages in Fig \ref{warp_Rvz}. As shown in each sub-figure, the vertical velocity increases with radial distance in different mono-age populations from 0 to 10 km s$^{-1}$ at 1 Gyr, for the others it is from 0 to 6$-$8 km s$^{-1}$, but for the oldest it is even less, maximum is around 5 km s$^{-1}$, they are definitely reflecting the warp signals. And the vertical velocities of clearly increase for most age bins are around 6$-$8 km s$^{-1}$ from 8 to 14 \,kpc except the last one, which is similar to \citet{Poggio2018} for the value of 5$-$6 km s$^{-1}$. Although there are some oscillations due to the Poissonian noise, as we will mention again in the next section 4, in order to get more points to show how the vertical velocity along with the distance and ensure we have enough data to do fitting, here we plot the velocity profile with the bins containing at least 20 stars. If we enlarge the bin size, the oscillation will be reduced. It is also clear to see that the younger age bins of top three panels have significantly smaller errors compared with the bottom three panels. Again, it might be caused by Poissonian noise. Moreover, the age accuracy is also becoming worse and worse as the stars are older and older. The youngest population vertical velocity is significantly larger than the oldest one in the top left and bottom right panel respectively, which might imply a warp amplitude difference. During this work, the size and the number of the bins are not constant for the different age sample, errors are very different in different populations. We have calculated the number to see their variation labeled in the Fig \ref{warp_Rvz}. There are many more young stars than old stars, which would explain that the poisson noise seems smaller at age  less than 5 \,Gyr. For the population at 9 \,Gyr, the number is very small so the large error bar could be caused by possion noise. The drop at large R for the last age bin might be caused by the Sun is not being on the line-of-nodes thus cause some stars not to move towards anticenter with different directions possibly so we think it is significant.

\section{Model} 

With the assumption that this vertical motion is contributed by warp, modeled as a set of circular rings that are rotated and whose orbit is in a plane with angle $i_w(R)$ with respect to the Galactic plane, as we can see many more details in \citet{Lop141}, the modeling process is displayed as follows:
\begin{equation}
V_{Z} = \Omega (R,z'=z-z_w)\sin [i_w(R)]\cos (\phi -\phi _w)+\dot{z_w}
\end{equation}
\begin{equation}
z_w(R,\phi )=R\tan [i_w(R)]\sin (\phi -\phi _w)
\end{equation}
where $\phi _w$ is the azimuth of the line of nodes, and $z_w$ is the height of the disk over the $b=0$ plane. We assume the greatest height of the warp to be 
\begin{equation}
z_w(R > R_\odot,\phi=\phi _w+\pi /2)\approx \gamma (R-R_\odot )^{\alpha }
\label{zwmax}
\end{equation}
and a variable line of nodes which has no extremely slow precession to do fitting, (i.e. $\dot{\phi _w}\ll \dot{\gamma }$) and  changed with the shape of the warp are adopted. To do so, we also assume a constant rotation speed $\Omega (R,z)=\Omega _{LSR}=238$ km/s; this may be slightly reduced for high $R$ or high $|z|$ \citep{Lop141}, but the order of magnitude does not change, and $V_Z$ is only weakly affected by a change of the rotation speed. Combining all these formulas and assumptions, we derive, for low angles $i_w(R)$, the low height disk warp model can be simplified reasonably as:
\begin{equation}
\label{model}
\begin{split}
V_{Z}(R > R_\odot)
\approx \frac{(R-R_\odot )^\alpha}{R}[\gamma \Omega_{LSR}\cos (\phi -\phi _w) \\
+\dot{\gamma }R\sin (\phi -\phi _w)]
\end{split}
\end{equation}

where $\phi _w$ is the azimuth of the line of nodes, $\gamma$  is the amplitude of the warp and $\dot{\gamma }$ describes the warp amplitude evolution, that is to say, $-$d($\gamma $)/d(age). We assume the exponent $\alpha =2$ ($kpc^{-1}$) \citep{Lop141} and set the line of nodes as $\phi _w$ ($deg$) as a free parameter, $\phi _w = variable$ deg. (in the literature the values are between -28, -5,  +15,  and 18 $deg$ \citep{Lop02b, Momany2006, Reyle2009, Skowron2019, Chen2019}. The constant $\alpha =1$(no units) \citep{Reyle2009} is also tested in our studies, and we checked that the conclusions derived from the fitting results are stable and not much affected by the value of $\alpha $. Here we just use the data with $R\ge 8.34$ \,kpc beyond the sun to get the best fitting value based on Markov chain Monte Carlo (MCMC) simulation provided by EMCEE \citep{Foreman$-$Mackey2013}. The model and method are used  maturely by our series of works in \citet{wang2020a} by fitting all the populations. With the help of the carefully selected sample, the present model and the popular MCMC method, we could obtain the likelihood distribution of the vertical velocity profile for fitting as:

\begin{equation}
\label{lnlike}
\begin{split}
\mathcal{L}(\{V_{\rm obs}(R_i|Z, Age)\}|\gamma,\dot\gamma, \phi_{w}) = \prod_i\exp\left[-{1\over{2}}\right.\\ 
(V_{\rm obs}(R_i|\left.Z, Age) - V_{\rm model}(R_i|Z, Age, \gamma, \dot\gamma, \phi_{w}))^2\right]
\end{split}
\end{equation}

During this likelihood, $R_i$ is the $i$th point of the fixed $Z$ grid in different age bins. It is emphasized here that each $R_i$ is naturally corresponding to a $\phi_i$ during the process and warp could vary with radius and azimuth angle, we use the information of Fig \ref{warp_Rvz} to constrain warp for this work. Please notice that we set the parameter as $\dot{\gamma}$ = $-$$d(\gamma)/d(age)$  by adopting a joint likelihood with 12 parameters to do simultanous fitting of all of age bins. In order to get the convergent parameters and save computer time, we just  choose a relatively smaller sampling size in our simulation, the MCMC size is 50*12*1000, and the step is 500. For a test, we also set the larger sampling size in MCMC, but the pattern is stable. As an attempt to explore the amplitude, line of nodes and maximum stellar warp height with age, our results are shown in the next section.

\section{Results} 

\subsection{Simplified analytical model fitting}

\begin{figure*}
  \centering
  \includegraphics[width=1.0\textwidth]{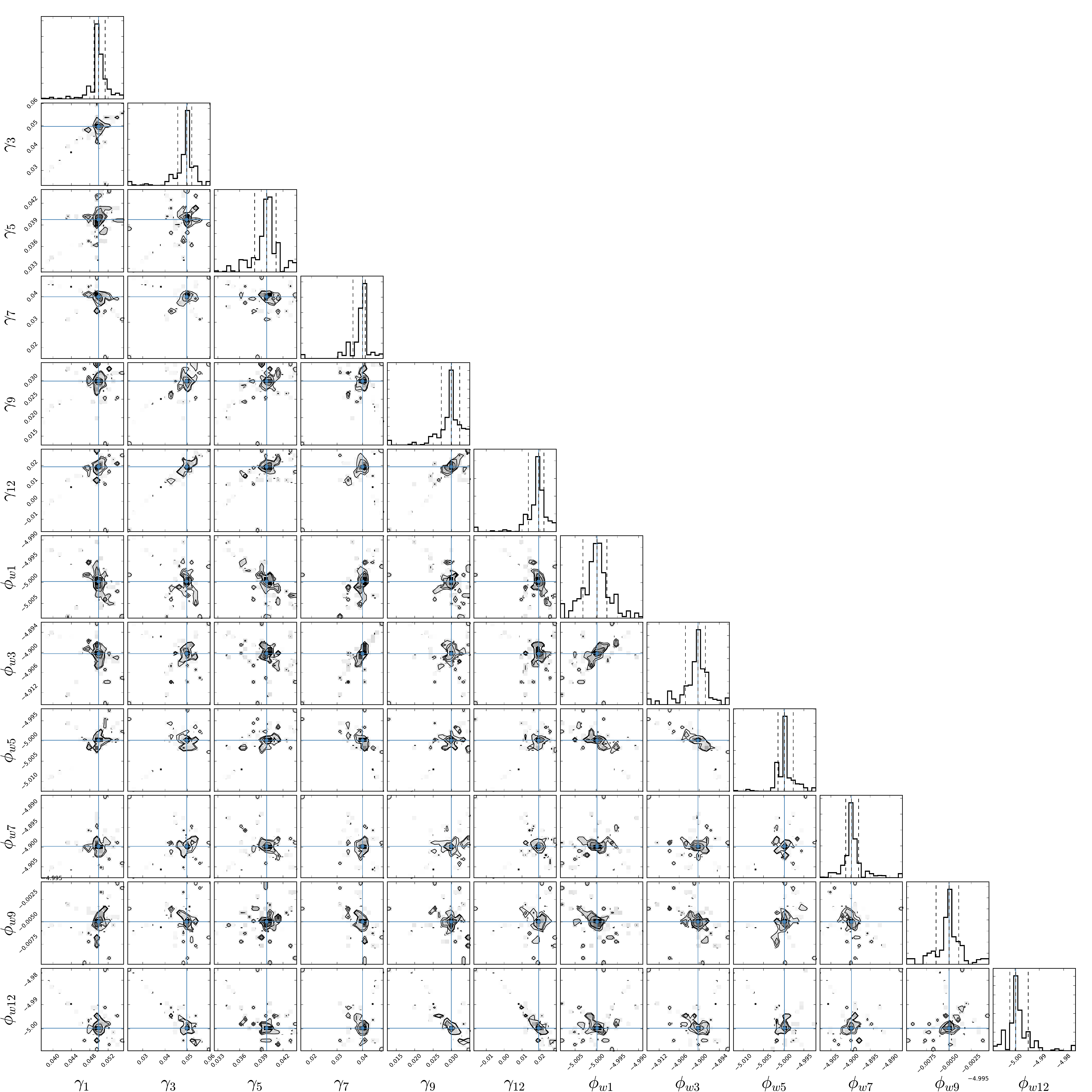}
  \caption{The likelihood distribution of the parameters ($\gamma $ and $\phi_{w}$) drawn from the MCMC simulation for 0-14 \,Gyr in the panel. The solid lines in the histogram panels indicate the maximum likelihood values of the parameters. The dashed lines indicate the 1$-\sigma$ regions defined by the covariance matrix. The panel is corresponding to the two parameters of each population in six bins simultaneously, namely, [1, 3 ,5, 7, 9, 12] \,Gyr, so there are twelve parameters.}
  \label{warpfitting81014}
\end{figure*}

It is well known and mentioned that the vertical bulk motions can be excited by the warp \citep{roskar10,Lop141,wang2018a, wang2020a}, which implies clearly vertical upward motions can be used to reveal the properties of warp, such as warp amplitude, precession rate and so on. The fitting results of the work in different age bins, by fitting simultaneously all the age bins, with $\alpha =2$ in model, are displayed in Fig \ref{warpfitting81014}. Some detailed warp features are revealed in the likelihood distribution of the parameters ($\gamma$  and $\phi_{w}$) drawn from the MCMC simulation in the next section.

\subsection{Warp parameter $\gamma$, $\dot{\gamma}$ and $\phi_{w}$ evolution with age}

\begin{figure*}
  \centering
  \includegraphics[width=1.0\textwidth]{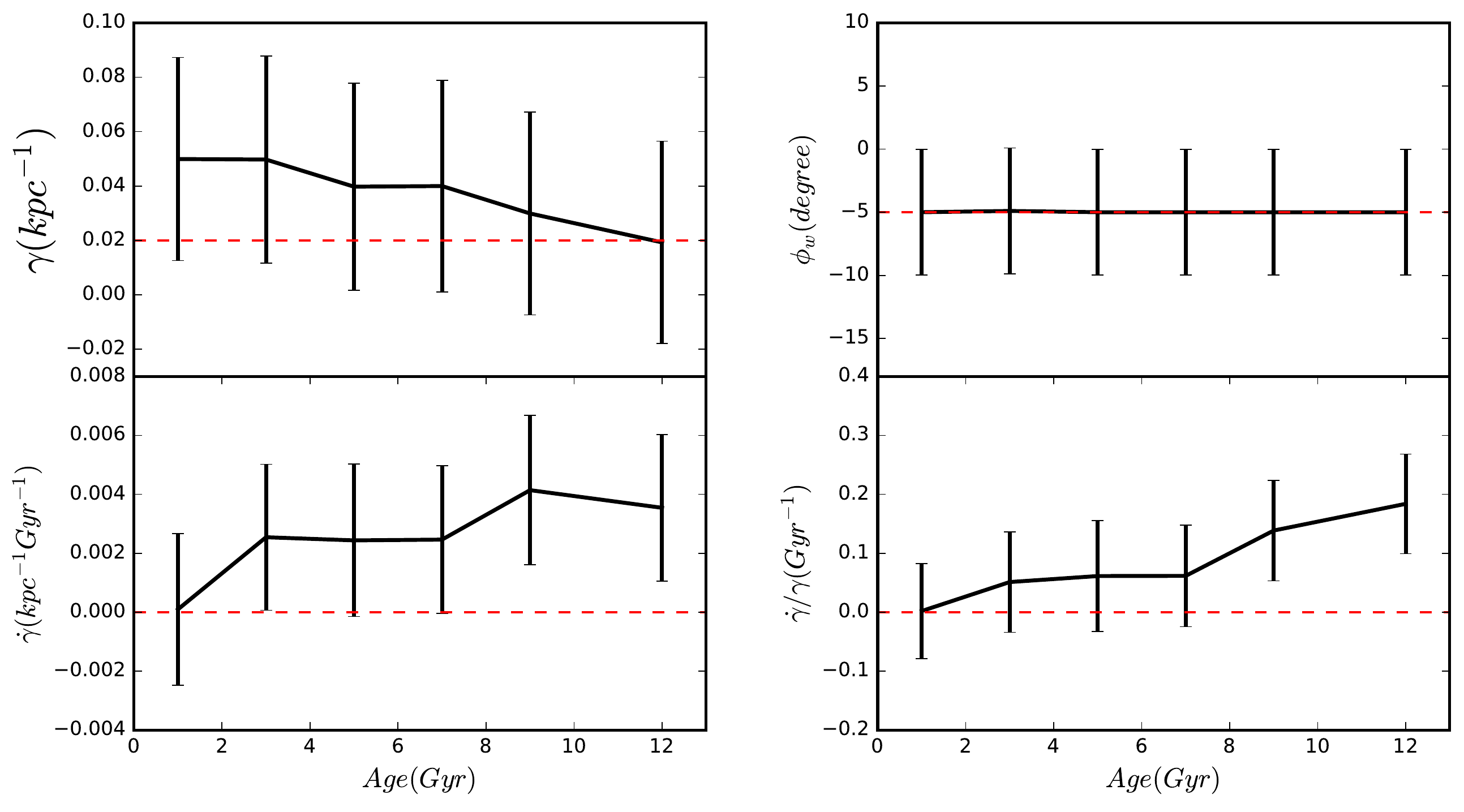}
  \caption{Top one shows the amplitude $\gamma$ ($kpc^{-1}$) evolution with age, the median value of six age bins are adopted in x axis: [1, 3 ,5, 7, 9, 12] \,Gyr; y axis is the MCMC values. Correspondingly, the bottom left panel is the $\dot{\gamma}$ ($kpc^{-1} Gyr^{-1}$) distribution for age with the error determined by bootstrap error. The top right panel is the stable azimuth of line of node along with age and the corresponding bottom right one is the natural product ($\dot{\gamma}$/$\gamma$) based on the left two figures, which is also increasing accompanied with errors, all these figures errors are from bootstrap process.}
  \label{warpevolution}
\end{figure*}

In Fig. \ref{warpevolution}, the amplitude $\gamma$ ($ kpc^{-1}$) evolution of warp with age is shown in the top left panel. We use the median value of six age bins, that is to say,  [1, 3 ,5, 7, 9, 12] \,Gyr, as the x axis value; and y axis value is the fitting results calculated by the MCMC according to its Gaussian distribution. The probability distribution and its peak are similar in all of the cases. The error is re-calculated again by bootstrap process. We can see that  there is variation of the $\gamma (kpc^{-1})$ with age with relatively large 1$-\sigma$ error, all these values are decreasing with age. Correspondingly, the bottom left panel is the warp amplitude derivative variation $\dot{\gamma}$ ($ kpc^{-1} Gyr^{-1}$) distribution with age, it has a variable increasing trend, these values are different from zero, implying that the warp is always existing but not a stationary structure ($\dot{\gamma}$ ($ kpc^{-1} Gyr^{-1}$) $\sim$ 0) and there is clear difference of populations existing. Moreover, there is also a stable feature for the azimuth of line of nodes in all populations, the value is almost fixed at about $-$5$^\circ$(degree) for the distribution displayed in the top right panel. The variations is very small due to the relative larger error makes it to be shown as a flatline, and some theoretical studies also support that the line of nodes is expected to be straight within R $\leq$ 4.5 disk scale lengths \citep{Shen2006, Joss2016}. The variation of $\phi_{w}$ due to precession is too small to be detected in all of the warp models are also mentioned in \citet{Lop02a, Dubinski2009, Jeon2009, Poggio2019}. Please note that there are only minor oscillations in the figure when we zoom in and there might be some intriguing physics in it.

We are considering the variation of $\phi_{w}$ with time negligible. We cannot distinguish in the vertical motion maps which is the dominant factor in $V_{Z}$, $\dot{\gamma}$ or $\dot{\phi_{w}}$, but the variation of $\phi_{w}$ due to precession is too small to be detected in all of the warp models \citep{Lop02a, Dubinski2009, Jeon2009}. Moreover, the fits depend on the model for warp shape. In \citet{Poggio2019}, they are also taking into account the variation of the amplitude of warp. They have got a long lived precessing model with large value by assuming a single value and no radial motions with the help of average of four geometrical simplified models. We suggest that the calculation of their precession with the very young population of \citet{Chen2019} is assuming it is similar for old and young population, also implied in Extended Data Fig. 3 for similar precession values of four models, and the model of warp they have used with Gaia DR2 is not good  because it does not reach high R, a good description of fits results of the old population in comparison with the young population warp is displayed in \citet{Zofia2020}, which is also showing younger populations like Cepheids are worse than whole populations. There is a possibility that when they get too high values of the vertical motions without precession, they have to introduce a too high precession to compensate it.  We are skeptical about the validity of their results that are needed to be investigated more.

As a natural product, the $\dot{\gamma}$/$\gamma$ is shown in the bottom right sub-figure, which is different from zero and increases with age. The figures derived from the MCMC simulations for all these $Age$ bins could give us a reliable estimation for these parameters of $\gamma$ ($kpc^{-1}$) and $\phi_{w}$ ($deg$) thanks to the peak of the maximum likelihood. Note that $\gamma$ has units of $kpc^{-1}$ for our adopted value of $\alpha$=2; it has no units when we set $\alpha$=1.

\subsection{$Z_{w}$: Greatest height of warp for younger ages}

\begin{figure}
  \centering
  \includegraphics[width=0.48\textwidth]{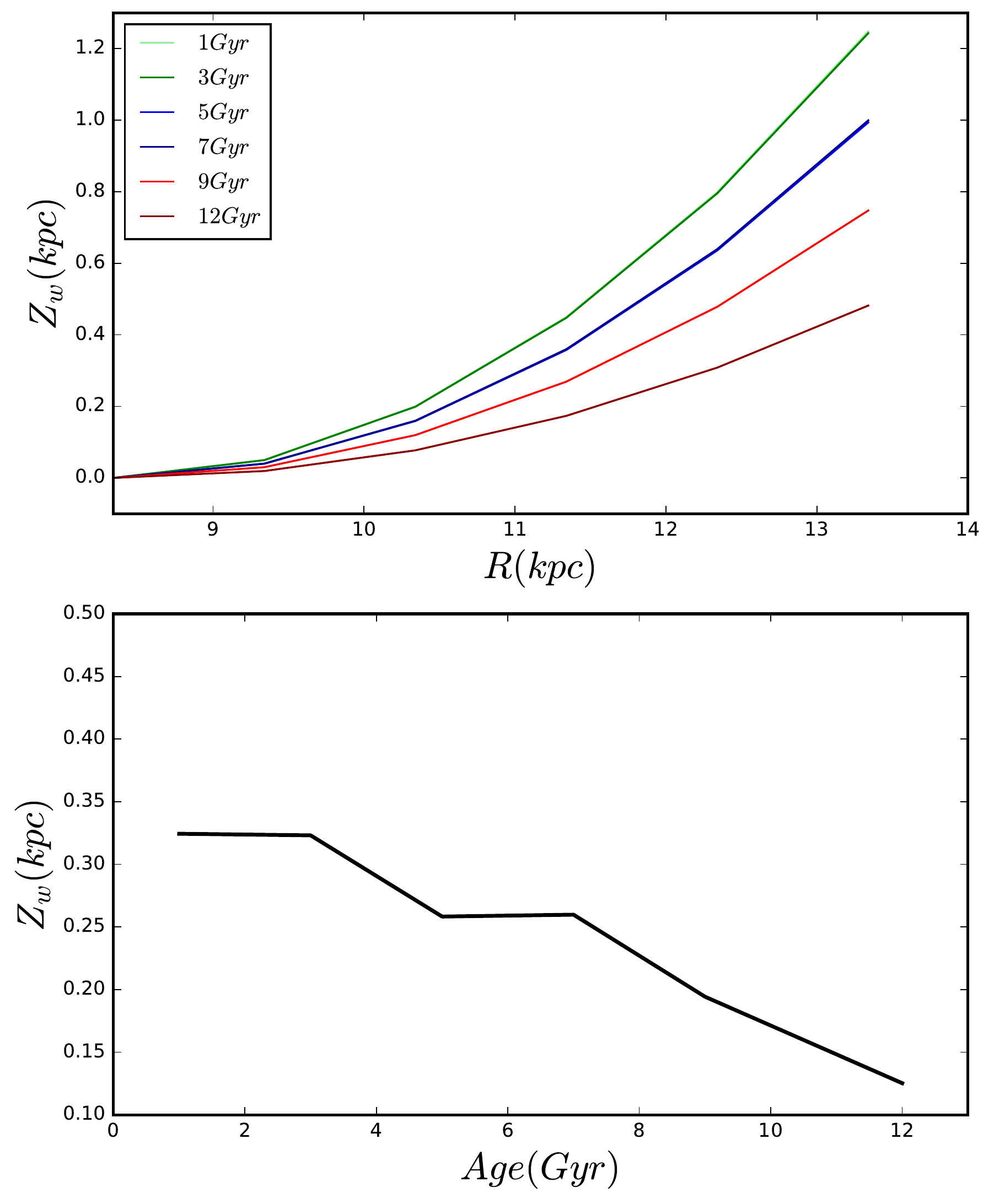}
  \caption{The top panel of the figure shows the greatest height of warp disk, defined in  Eq. (\ref{zwmax})  with some assumptions, distributions with Galactic radial distance, in which the different color solid lines represent different ages of  [1, 3 ,5, 7, 9, 12] \,Gyr. Please notice that there are six age bins but two lines are overlapped each other (1 and 3, 5 and 7). It has positive gradient and increased with $R$ in different populations. The bottom one is the median value of the $Z_{w}$ with age, a clear decreasing are here.}
  \label{ZwageR}
\end{figure}

In Fig. \ref{ZwageR}, we show the distribution of the greatest height of warp disk with different age bins (bottom panel) and distance of different populations (top panel). The top panel suggests that there is an increasing trend for the height along with distance in all age bins. We also see the younger populations are higher than the old ones. Note that there are six age bins but two lines are overlapped. For the bottom one, we can clearly see there is clear decreasing pattern for all median heights in all age populations. It is consistent with the results in Fig. \ref{warpevolution}, meaning that the warp appears to be a long lived not-stationary structure and, more importantly, there is a clear difference for different populations again.

\subsection{The comparison of Model and Sample}

\begin{figure*}
  \centering
  \includegraphics[width=0.98\textwidth]{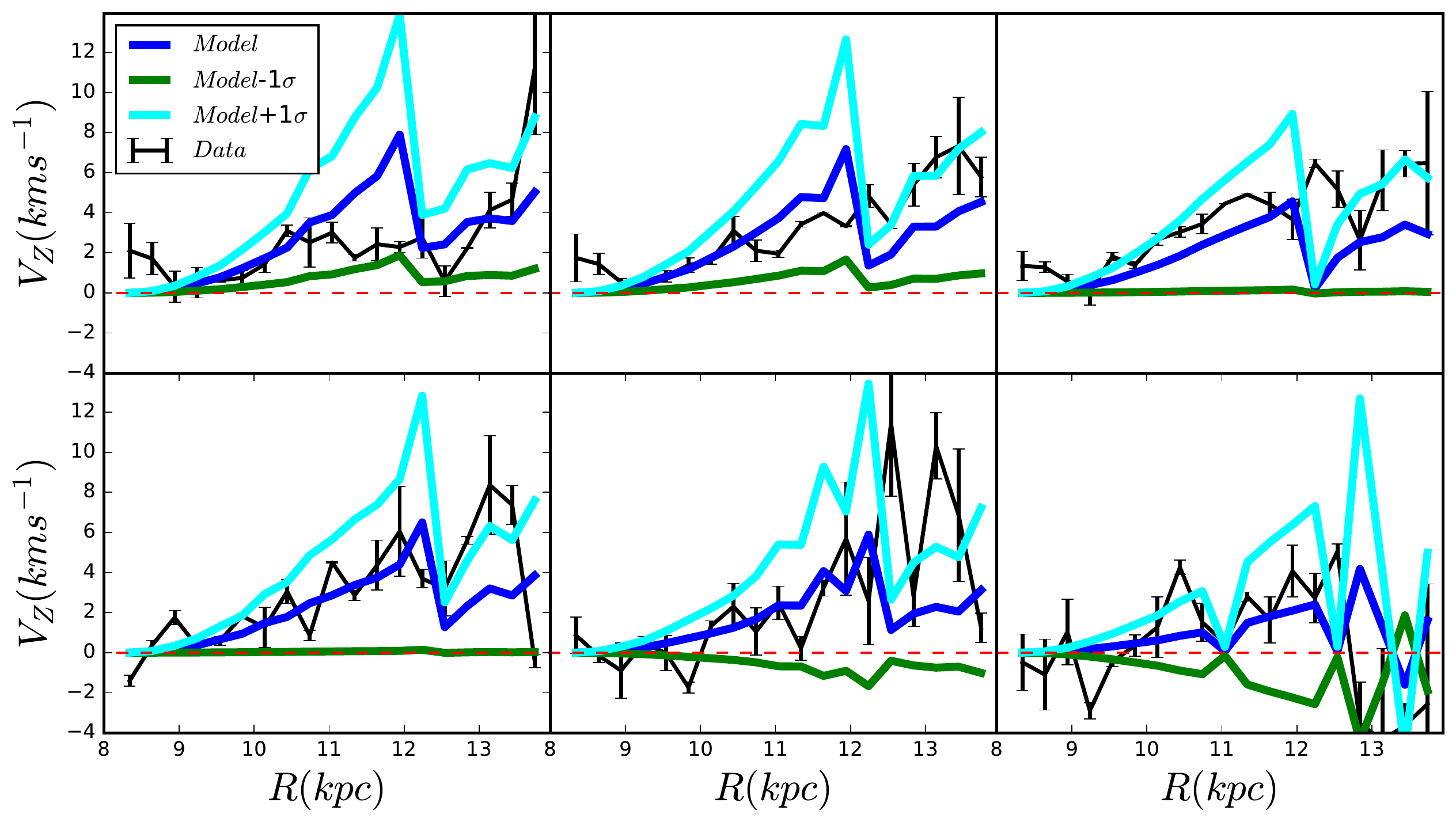}
  \caption{Model and data comparison in six age populations during this work, from the top left one to the bottom right one is corresponding to median value of [1, 3 ,5, 7, 9, 12] \,Gyr: the black curve is the observed vertical velocity distribution with Galacto-centric radial distance, the blue one is the model profile, the cyan one is the model profile plus 1$\sigma$ and the green is model profile plus $-$1$\sigma$. The oscillations in the model are due to the fact that the model with sine and cosine function and different values of $R$ have different ranges of $\phi$ for most stars.}
  \label{warp_modeldatafitting}
\end{figure*}

In order to check our robustness of the derived results, we have finished the comparison of Model of Eq. (\ref{model}) and data in six age populations during this work as shown in Fig. \ref{warp_modeldatafitting}. From the top left panel to the bottom right one, the median values of $V_Z$ for [1, 3 ,5, 7, 9, 12] \,Gyr populations are plotted. The black bold solid line with error bars is the observed vertical velocity distribution with Galactocentric radial distance in each panel; the blue one is the model profile with the Monte-Carlo fitting, the cyan one is the model profile plus 1$\sigma$ and the green is model profile plus $-$1$\sigma$. As we can see here, for the general trend, the matching results of all populations between the data and model are good within the uncertainties. The goodness of fit is not good enough at the left boundary in some populations. However, again, the general trend is quite good in 1$\sigma$ for most of the points. Few mismatches shown here are caused by some reasons such as, in some regions, our model is a simple model, it is not very perfect and suitable for us to describe different populations in some cases; when the distance is larger, the error of stellar parameter and age precision is becoming worse, making it cannot be fitted quite well; the Poisson noise due to the the number of our sub-sample is not large enough; the Sun is not being on the line-of-nodes thus cause some stars not to move towards anticenter with different directions; extinctions in some regions are not perfect. 

In order to try to test the Poissonian noise to mislead us for the conclusion, we just compare our model with data consisted of at least 100 stars in each bin, which has small differences for the pattern shown in Fig. \ref{warp_modeldatafitting} and  Fig. \ref{warp_Rvz}. The overall trend is better except fewer boundary points deviate, which can not change our conclusion at all. So we suggest all features mentioned in the previous paragraphs are robust and intriguing. It seems that there is systematical peak in the model at R=12 \,kpc of Fig. \ref{warp_modeldatafitting}, we actually have a test for it to suggest that our model with sine and cosine function and different values of R have different ranges $\phi$ of for most stars. If we have a simple Monte Carlo simulation, there are some clear oscillations and peaks located around 12 \,kpc, so it is expected and possibly caused by our model distribution properties.

Therefore, in short summary, we think that these stable features are real and these observational evidences strongly support that the warp is a long lived and not-steady S shape structure, and it also implies the warp evolution is relatively uniform and there is a clear difference for different age bins. So far,  we only have a few points span from 8$-$14 \,kpc with relatively large and not-perfect age error, it would be worth for us to further investigate in more details for this structure with the help of sample consisted of larger distance range, more stars, and more accurate age. Some scenarios are given in the next section.

\section{Discussions}

Vertical non-axisymmetries  and wave-like density patterns are found in the solar neighborhood and in the outer disk \citep{Widrow12, Williams13, Xu15, wang2018a, wang2018b, wang2018c, Carlin13,  Carrillo17,Carrillo2019, Pearl17}. As mentioned  and implied in \citet{wang2020a}, many mechanisms including warp might be coupled together under the same complexed distribution function to cause the complicated Galacto-seismology signals. Scenarios for producing these structures such as warp dynamics, minor mergers or interactions with nearby dwarfs or satellites \citep{Gomez13, Laporte18, Donghia16, Laporte2019} and the effects of even lower-mass dark matter sub-halos have also been invoked as a possible explanation \citep{Widrow14}. Notice that vertical velocity asymmetry can be applied to constrain the warp signal and it is acceptable that we use the vertical motions to acquire the warp amplitude and its variation, although other causes different from warp may contribute to the vertical motions too.
 
The kinematical features of the Galactic warp around its line-of-nodes located close to the Galactic anti-center region are discussed in \citet{Liu173}, where it was found that the vertical bulk motion of younger red clump stars are significantly larger than that of the older ones, which is consistent with our results. We have got a not-steady warp. A variation of warp amplitude with stellar population age is in principle against a steady warp due to steady gravitational forces and it is more in favour of the models in which gas is necessary for warp formation, or similar scenarios. The reason is that  the young population, tracing the gas, will always have larger warp amplitude. 

According to \citet{Skowron2019b}, there are mainly two classes of warp formation mechanisms. One is that the warp formed by the gravitational interactions, for example, with satellite galaxies or a misaligned dark matter halo. The other one is non-gravitational mechanisms, e.g, accretion of intergalactic gas onto the disc\citep{Lop02a}, or interactions with intergalactic magnetic fields. Non-gravitational mechanisms such as magnetic fields or hydrodynamical pressure from infalling gas would act on the gas and only affect the young stars \citep{Guijarro2010, Sellwood 2013}, thus we should see all signals of younger ones are stronger than the old ones, and we do get that clearly. Therefore, we think the gravitational scenario should not be the reason at least for this tracer.
 
Young population traces the gas, whereas old population had more time to reduce the amplitude of the warp due to the self-gravity in the models in which the torque affects mainly to the gas and not to the stars. Our current results support the warp might be contributed by the non-gravitational interaction models, which do not agree with the viewpoints of \citet{Poggio2018} by using upper main sequence stars and giants as two age populations.
 
An age dependence both on position and kinematics of the Galactic warp has been observed by \citet{Amo17,Gomez2019}, and confirmed for our results in Fig. \ref{warp_Rvz} and Fig. \ref{ZwageR}. We could see there are some vertical velocity differences and the greatest height difference in different age bins clearly. \citet{Amo17} thought that the warp shapes and amplitude of northern part are stronger than the southern part and the north$-$south asymmetry was also presented in \citet{Reyle2009, Momany2006}, which we can not test here, the reason is that LAMOST mainly cover the Galactic Anti-Center in the north, we don’t have many stars in the south sky, it is impossible to discuss this difference, but we plan to work out it in the future. \citet{Gomez2019} showed the amplitude of OB stars corresponding to younger populations is weaker than the RGB stars corresponding to older ones by calculating the onset radius and height of warp, and thus suggesting that warping disk of older population is more pronounced or stronger, which is not consistent with our results implying that the warp amplitude is variable and decreasing in different age bins. The discrepancy might be due to the fact that we use the direct age bins and greatest height to describe the warp amplitude, but \citet{Gomez2019} use the height and indirect age indicators with OB and RGB stars, so the methods, assumptions and population effects will be important for the discrepancy. Furthermore, \citet{Skowron2019} showed that their results, by using Cepheids similar with OB stars, are in contradiction to \citet{Gomez2019} with that the Cepheids warp height is similar with RGB stars.
 
We also notice that the recent results of \citet{Chen2019} about the warp for very young populations give the highest amplitude of the warp so far for stellar populations, which also supports our conclusion that younger ones are stronger ones. A young population warp larger than the old population one has also been demonstrated by \citet{Zofia2020} using Gaia-DR2 density maps extended up to $R=20$ kpc, thanks to the use of deconvolution techniques of parallax errors. In our series of works shown by \citet{wang2020a}, we already got the conclusion that the warp is a long lived and not steady structure by adopting all populations and fix the line of node at 5 \,deg to do the Monte Carlo fitting. It was a relatively simple investigation, but the conclusion was similar to the present work.

In short, to some extent, our results and implications are similar to those obtained by \citet{Reyle2009, Amo17, Chen2019, Liu173, Zofia2020} and others, although we have some difference with \citet{Gomez2019, Poggio2018, Skowron2019} with some possible reasons. Further works need to be done to clarify these differences.
 
In \citet{Chen2019}, their line of nodes is around 18 $deg$ with the help of LONs in different radial bins. In other literature the values are between -5 and +15 $deg$ \citep{Lop02b, Li2020, Reyle2009, Momany2006}. \citet{Skowron2019} gave the larger negative value of -28 $deg$. Our results are different from some works corresponding larger values of the warp line of node, but it is well consistent with the value of 5 $\pm$ 10 deg used by \citet{Lop141} and supporting strongly the works finished by \citet{Lop02b}. It is intriguing to report here that the red clump stars with 2MASS finished by \citet{Lop02b} yielded $\phi_{w}= $-$5\pm 5 \deg$. 

\citet{Li2020} also get a different value around 12 $deg$ by using \citet{Poggio2017} simplified model, the difference is caused by that the models and assumptions have much differences, but \citet{Li2020} find that the warp signal of thick disk population is weaker than the thin disk, which is consistent with our main conclusion.

A three-dimensional map of the Milky Way with the help of classical Cepheid variable stars and the simple model of star formation in the spiral arms was used to reveal the shape of the young stellar disk in \citet{Skowron2019}. It also mapped the distribution of Cepheid tracers with age for which the range is within a few hundred $Myr$, much smaller than the range of ages in our sample. Therefore, this is the first time here the warp evolution is followed with complete age sampling of the Milky Way.
 
In the future, we will go farther than 20$-$25 \,kpc of the disk with the state of the art of warp model and more accurate and  larger sample. We think we can constrain our galaxy warp better and better, as we mentioned, since there are still relatively large errors in our results. We could also compare stellar warp with gas disk and dust disk warp signals. Moreover, we could use $[Fe/H]$ and $[\alpha/Fe]$ as population indicator to see more evolution features of the warp. For the target of this work, we just explore the warp variation with age as an attempt by using a simplified model.

\section{Conclusion} 

In this paper, using LAMOST$-$Gaia combined red clump giant stars with unprecedented proper motion and age accuracy, we investigate the evolution of warp amplitude, line of nodes, the greatest height and its variation with age. The greatest height of the warp is decreasing with the age and increasing with distance in mono-age populations: the younger populations are strongly warped than the old ones. And we also observe the amplitude's temporal evolution and its first derivative with time have a decreasing and increasing pattern respectively. A stable azimuth of line-of-node is $-$5 degree  is shown in this work. Our results are similar to some of recent works, but we use the standard candles with age to quantify warp amplitude evolution.

All these observational results are supporting the warp is not a transient structure, and also implying strongly that warp evolution is non-uniform, long lived, non-steady structure.
We conclude that the warp might be induced by the non-gravitational interaction scenarios: gas infall onto the disc\citep{Lop02b} or magnetic fields \citep{Bat98} or similar classes. It might reflect some puzzling evolution of the warp that should be further studied. 

Both the simple model and data used here can be improved. We need a better warp model and more accurate age measurements to further research this S-like stellar disk. Our work might be of vital importance for us to investigate more properties of the warp and  more work will be shown in our series of works.

 \acknowledgements
We would like to thank the anonymous referee for his/her helpful comments. HFW is supported by the LAMOST Fellow project, National Key Basic R\&D Program of China via 2019YFA0405500 and funded by China Postdoctoral Science Foundation via grant 2019M653504, Yunnan province postdoctoral Directed culture Fundation and the Cultivation Project for LAMOST Scientific Payoff and Research Achievement of CAMS-CAS. Y.H. acknowledges the National Natural Science Foundation of China U1531244,11833006, 11811530289, U1731108, 11803029, and 11903027 and the Yunnan University grant No.C176220100006 and C176220100007. MLC was supported by grant PGC-2018-102249-B-100 of the Spanish Ministry of Economy and Competitiveness. JLC acknowledges support from the U.S. National Science Foundation via grant AST-1816196. HWZ is supported by the National Natural Science Foundation of China under grant number 11973001 and supported by National Key R\&D Program of China No. 2019YFA0405500. HFW is fighting for the plan ``Mapping the Milky Way Disk Population Structures and Galactoseismology (MWDPSG) with large sky surveys" in order to establish a theoretical framework in the future to unify the global picture of the disk structures and origins with a possible comprehensive distribution function. We pay our respects to elders, colleagues and others for comments and suggestions, thanks all of them. The Guo Shou Jing Telescope (the Large Sky Area Multi-Object Fiber Spectroscopic Telescope, LAMOST) is a National Major Scientific Project built by the Chinese Academy of Sciences. Funding for the project has been provided by the National Development and Reform Commission. LAMOST is operated and managed by National Astronomical Observatories, Chinese Academy of Sciences. This work has also made use of data from the European Space Agency (ESA) mission {\it Gaia} (\url{https://www.cosmos.esa.int/gaia}), processed by the {\it Gaia} Data Processing and Analysis Consortium (DPAC, \url{https://www.cosmos.esa.int/web/gaia/dpac/consortium}). Funding for the DPAC has been provided by national institutions, in particular the institutions participating in the {\it Gaia} Multilateral Agreement.

\end{document}